\documentclass[12pt,journal,onecolumn]{IEEEtran}

\usepackage{threeparttable,placeins,makecell,stfloats,cite}
\usepackage{threeparttable,arydshln}
\usepackage{amsmath,amssymb,bm}
\usepackage[dvips]{graphicx}
\usepackage[usenames]{color}
\usepackage{amsfonts}
\usepackage{latexsym}
\usepackage{multirow}
\usepackage{caption}
\usepackage{rotating}
\usepackage{float}
\usepackage[format=hang]{subfig}
\usepackage{url}
\usepackage{cite}

\begin{document}

\title{Design of Power-Imbalanced SCMA Codebook}

\author{Xudong~Li, Zhicheng~Gao, Yiming~Gui, Zilong~Liu, Pei~Xiao, Lisu~Yu
\thanks{
	X. Li and Z. Gao are with the Laboratory of Security Insurance of Cyberspace, School
	of Science, Xihua University, Chengdu 610039, China (e-mail: lixudong73@163.com, 1447152594@qq.com). Y. Gui is with the School of Economics, Xihua University, Chengdu 610039, China (e-mail: ymgui\_18@163.com). Z. Liu (\textit{Corresponding Author}) is with School of Computer Science and Electronics Engineering, University of Essex, UK (e-mail: zilong.liu@essex.ac.uk). P. Xiao is with the Institute for Communications, 5G Innovation Centre, University of Surrey, UK (e-mail: p.xiao@surrey.ac.uk). L. Yu is with the School of Information Engineering, Nanchang University, Nanchang 330031, China, also with the State Key Laboratory of Computer Architecture, Institute of Computing Technology, Chinese Academy of Sciences, Beijing 100190, China (e-mail: lisuyu@ncu.edu.cn). }
}

\maketitle

\begin{abstract}
Sparse code multiple access (SCMA) is a promising multiuser communication technique for the enabling of future massive machine-type networks. Unlike existing codebook design schemes assuming uniform power allocation, we present a novel class of SCMA codebooks which display power imbalance among different users for downlink transmission. Based on the Star-QAM mother constellation structure and with the aid of genetic algorithm, we optimize the minimum Euclidean distance (MED) and the minimum product distance (MPD) of the proposed codebooks. Numerical simulation results show that our proposed codebooks lead to significantly improved error rate performances over Gaussian channels and Rayleigh fading channels.
\end{abstract}


\section{Introduction}
Non-orthogonal multiple access (NOMA) is a disruptive multiuser communication technique for the 5G networks and beyond. In recent years, NOMA has attracted increasing research attention owing to its great potential of meeting various stringent quality-of-service requirements such as considerably higher spectrum efficiency, massive connectivity, and ultra-low access latency \cite{Liu2017,Liu2021}. Inspired by low density signature based code-division multiple access (LDS-CDMA) \cite{Hoshyar2008}, sparse code multiple access (SCMA) has emerged as a competitive code-domain NOMA scheme \cite{Nikopour2013}. In SCMA, the multiuser detection (MUD) is conducted by efficiently exploiting the sparsity of codebooks with message passing algorithm (MPA). Because of this, the error rate performances of SCMA approach to that of the maximum likelihood receiver. Moreover, SCMA signals enjoy the so called ``constellation shaping gain" by mapping the input bits to properly designed multiple multidimensional codewords.

Codebook design is of paramount importance for enhanced bit error rate (BER) performances in SCMA systems. Numerous codebook constructions have been proposed \cite{Taherzadeh2014,Bao2016,Zhang2016,Lim2018,Mheich2018,Yu2018,Cai2016,Zhou2017,Li2018}, albeit its optimal design remains largely open. Most existing codebook design schemes start from a multidimensional mother constellation (MC) with large minimum Euclidean distance (MED) or large minimum product distance (MPD) \cite{Boutros1996}. Subsequently, multiple sparse codebooks  are generated through spreading matrix and a series of constellation operations such as interleaving, permutation, and phase rotation. Despite extensive studies in the literature, the state-of-the-art codebook design is carried out channel by channel \cite{Chen2020}. That is, an optimized codebook with respect to one channel may not perform well in another.

In this paper, building upon our preliminary results in \cite{Liu2019}, we develop a novel class of power-imbalanced SCMA codebooks having power variation among different users for downlink transmission.  By leveraging the power-imbalance (a unique trait of power-domain NOMA), we aim to take advantage of the ``near-far effect" among different users in order to attain enhanced MPA decoding. Under the generic Star-QAM MC structure \cite{Yu2018}, we first optimize the user-wise MPD\footnote{In high SNR region, as most transmission errors occur in one user, the user-wise MPD dominates the SCMA error rate performance.} analytically and then adopt genetic algorithm (GA) to optimize the MPD of the resultant superimposed constellation (inspired by \cite{Klimentyev2017}), where GA is a stochastic optimization approach which has found wide applications in engineering. Numerical simulation results show that our proposed codebooks lead to significantly improved bit error rates (BERs) over both Gaussian channels and Rayleigh fading channels.

The rest of the paper is organized as follows. Section II introduces the SCMA system model. Section III proposes a new class of power-imbalanced SCMA codebooks, followed by its optimization with the aid of GA. Comparison between different codebooks and BER simulations are presented in Section IV. Section V concludes the paper.

\section{SCMA System Model}
Let us consider a  downlink SCMA system, in which the base station transmitter simultaneously communicates with $J$ users sharing $K$ resources nodes. In this system, we aim to attain $J>K$ with the overloading factor of $\lambda={J}/{K}>1$. Unlike LDS, an SCMA encoder directly maps the incoming $\log_2M$ bits $\bm{b}$ to a $K$-dimensional complex constellation $\bm{\chi}$ which has size of $M$ and then superimpose the codewords of the $J$ users for transmission. The received signal vector $\bm{y}=(y_1,y_2,\cdots,y_K)^T$ at one user end can be expressed as
\begin{equation}
	\bm{y}=\text{diag}(\bm{h})\sum_{j=1}^{J}\bm{x}_j+\bm{n},
	\label{receive_signal}
\end{equation}
where $\bm{h}=(h_1,h_2,\cdots,h_K)^T$ denotes the channel vector with $h_k\sim\mathcal{CN}(0,1)$, and $\text{diag}(\cdot)$ stands for the vector diagonalization. $\bm{x}_j=(x_{1,j},x_{2,j},\cdots,x_{K,j})^T$ refers to a $K$-dimensional complex codeword of user $j$, where $1\leq j \leq J$, which is selected from the codebook $\bm{\chi}_j$. $\bm{n}=(n_1,n_2,\cdots,n_K)^T$ denotes the additive white Gaussian noise (AWGN) vector with zero mean and $\sigma^2\bm{I}_K$ variance.

A sub-optimal SCMA codebook design can be summarized as the following two steps: first, generate a multidimensional mother constellation $\bm{C}\in \mathbb{C}^{N\times M}$ having good error rate performances in certain channels, where $N$ denotes the number of non-zero entries in each $K$-dimensional sparse codeword with $N<K$. Then, construct multiple codebooks through specific constellation operation matrix $\bm{\Delta}\in \mathbb{C}^{N\times N}$ and spreading matrix $\bm{V}\in \mathbb{B}^{K\times N}$. For user $j$, its codebook can be expressed as $\bm{\chi}_j=\bm{V}_j\bm{\Delta}_j\bm{C}_{MC}$, where $1\leq j\leq J$.

Similar to LDS, specific constellation operation matrix $\bm{\Delta}$ and spreading matrix $\bm{V}$ can be combined together and represented as signature matrix $\bm{S}_{K\times J}=[\bm{S}_{K\times  1}^{1},\cdots,\bm{S}_{K\times  1}^{J}]$, where $\bm{S}_{K\times 1}^{j}=\bm{V}_j\bm{\Delta}_j\bm{I}_c$ and $\bm{I}_c$ denotes a column vector of $N$ 1's. For user $j$, its codebook can also be expressed as
\begin{equation}
\bm{\chi}_j=ezc\left(\text{diag}\left(\bm{S}_{K\times 1}^j\right)\right)\bm{C}_{MC},
\end{equation}
where $ezc\left(\text{diag}\left(\bm{S}_{K\times 1}^j\right)\right)$ denotes the resultant matrix after removing the all-zero columns.

As an example, suppose the spreading matrix and constellation operation matrix of user $j$ are $\bm{V}_j=\left[\begin{matrix}
0 & 1 & 0 & 0\\
0 & 0 & 0 & 1\\
\end{matrix}\right]^{T}$ and $\bm{\Delta}_j=\left[\begin{matrix}
e^{j\varphi_1} & 0 \\
0 & e^{j\varphi_2}
\end{matrix}\right]$, respectively. Then, we have $\bm{I}_c=\left[1,1\right]^{T}$ and $\bm{S}_{K\times 1}^j=\bm{V}_j\bm{\Delta}_j\bm{I}_c=\left[
0,e^{j\varphi_1},0,e^{j\varphi_2} \right]^{T}$. After obtaining the signature matrix, we can further calculate $ezc\left(\text{diag}\left(\bm{S}_{K\times 1}^j\right)\right)$ in \eqref{ezc} to verify that $ezc\left(\text{diag}\left(\bm{S}_{K\times 1}^j\right)\right)=\bm{V}_j\bm{\Delta}_j$.
\begin{equation}
\label{ezc}
\begin{array}{ccc}
\left[\begin{array}{cccc}
0&0&0&0\\
0&e^{j\varphi_1}&0&0\\
0&0&0&0\\
0&0&0&e^{j\varphi_2}
\end{array}\right] & \rightarrow &
\left[\begin{matrix}
0 & 0 \\
e^{j\varphi_1} & 0 \\
0 & 0 \\
0 & e^{j\varphi_2} \\
\end{matrix} \right]
\end{array}.
\end{equation}

Due to the sparsity of codebook, there are only $d_f$ users superimposed over each resource node, where $d_f<J$. The sparsity of an SCMA system can be seen from its corresponding signature matrix $\bm{S}_{K\times J}$. It is noted that User $j$ has active transmission over resource node $k$ if and only if $S_{k,j}\neq0 (1\leq k \leq K, 1\leq j \leq J)$. Different from the Latin structure considered in \cite{Deka2020,Xiao2018}, in this paper, we conceive the following two signature matrices $\bm{S}_{4\times6}$ and $\bm{S}_{5\times10}$

\begin{equation}
\label{factor graph4_6}
\bm{S}_{4\times6}=\left[\begin{matrix}
0 & z_1 & z_2 & 0 & z_3 & 0 \\
z_1 & 0 & z_2 & 0 & 0 & z_3 \\
0 & z_3 & 0 & z_2 & 0 & z_1 \\
z_3 & 0 & 0 & z_2 & z_1 & 0 \\
\end{matrix} \right].
\end{equation}

\begin{equation}
\label{factor graph5_10}
\bm{S}_{5\times10}=\left[\begin{matrix}
z_1 & z_2 & z_3 & z_4 &  0  &  0  &  0  &  0  &  0  &  0  \\
z_4 &  0  &  0  &  0  & z_1 & z_2 & z_3 &  0  &  0  &  0  \\
0  & z_3 &  0  &  0  & z_4 &  0  &  0  & z_1 & z_2 &  0  \\
0  &  0  & z_2 &  0  &  0  & z_3 &  0  & z_4 &  0  & z_1 \\
0  &  0  &  0  & z_1 &  0  &  0  & z_2 &  0  & z_3 & z_4 \\
\end{matrix} \right].
\end{equation}

By replacing all the nonzero values in $\bm{S}_{4\times6}$ and $\bm{S}_{5\times10}$ by ones, we can obtain two indicator matrices of orders $4\times 6$ and $5\times 10$, respectively. The sparsity may be ensured if the factor graph associated to each indicator matrix has the minimum cycle of 6, where the cycle in a factor graph is formed by joining those adjacent edges across several user nodes and resource nodes \cite{Liu2021}. Moreover, the inherent sparsity of SCMA signals is exploited at the receiver by the MPA detector with complexity of $O(JNM^{d_f})$. Compared with the maximum likelihood detection algorithm, the decoding complexity of MPA detector is drastically reduced.

\section{Proposed Power-Imbalanced Codebook Design}
MC plays an important role in SCMA codebook design. In this paper, we employ the Star-QAM MC structure proposed in \cite{Yu2018}. An example of 2-dimensional MC is shown below:
\begin{equation}
\setlength{\arraycolsep}{1.2pt}
\label{MC}
\bm{C}_{2\times M}=\left[\begin{matrix}
\omega_{\frac{M}{2}}  & \omega_{\frac{M}{2}-1}  & \cdots  & \omega_{1} & -\omega_{1} & \cdots & -\omega_{\frac{M}{2}-1} & -\omega_{\frac{M}{2}}\\
-\omega_{1}  & \omega_{2}  & \cdots  & \omega_{\frac{M}{2}} & -\omega_{\frac{M}{2}} &  \cdots & -\omega_{2} & \omega_{1}\\
\end{matrix} \right],
\end{equation}
where $\omega_i=(i-1)(\omega-1)+1$ and $\omega$ is a positive scalar parameter to be optimized, where $\omega>1$. Each dimension of the MC has energy of
\begin{equation}
	E=\frac{M(M-1)(M-2)(\omega-1)^2}{12}+\frac{M(M-2)(\omega-1)}{2}+M.
\end{equation}

Aimed at increasing the power diversity of codebook, the nonzero elements of the corresponding signature matrix $\bm{S}_{K\times J}$ can be specified as $z_i=\sqrt{E_i/E}e^{j\varphi_i}$, where $1\leq i \leq d_f$ and $\varphi_i$ denotes the phase rotation angle of $z_i$. The dimensional energy of codebook can be calculated as  $\Vert[\omega_{\frac{M}{2}},\cdots,\omega_{1},-\omega_{1},\cdots,-\omega_{\frac{M}{2}}]\cdot z_i\Vert^2=E\cdot \left(E_i/E\right)=E_i$, and the corresponding dimensional energy matrices $\bm{E}_{4\times 6}$ and $\bm{E}_{5\times 10}$ can be written as

\begin{equation}
\label{energy_matrix_F4x6}
\bm{E}_{4\times6}=\left[\begin{matrix}
0 & E_1 & E_2 & 0 & E_3 & 0 \\
E_1 & 0 & E_2 & 0 & 0 & E_3 \\
0 & E_3 & 0 & E_2 & 0 & E_1 \\
E_3 & 0 & 0 & E_2 & E_1 & 0 \\
\end{matrix} \right].
\end{equation}

\begin{equation}
\label{energy_matrix_F5x10}
\bm{E}_{5\times10}=\left[\begin{matrix}
E_1 & E_2 & E_3 & E_4 &  0  &  0  &  0  &  0  &  0  &  0  \\
E_4 &  0  &  0  &  0  & E_1 & E_2 & E_3 &  0  &  0  &  0  \\
0  & E_3 &  0  &  0  & E_4 &  0  &  0  & E_1 & E_2 &  0  \\
0  &  0  & E_2 &  0  &  0  & E_3 &  0  & E_4 &  0  & E_1 \\
0  &  0  &  0  & E_1 &  0  &  0  & E_2 &  0  & E_3 & E_4 \\
\end{matrix} \right].
\end{equation}

There is significant power diversity, row-wise or column wise,  with $E_i\neq E_j, \forall 1\leq i<j\leq d_f$. The power imbalance among different users is introduced by $\bm{E}_{4\times 6}$  with $E_1+E_3\neq 2E_2$ and $\bm{E}_{5\times 10}$  with $E_1+E_4\neq E_2+E_3$.

After obtaining the new signature matrix $\bm{S}_{4\times6}$, as an example, the  codebook of User 1 can be expressed as
\begin{equation}
	{{\mathbf{\bm{\chi }}}_{1}}=\left[\begin{matrix}
	0 & 0 & 0 & 0  \\
	\omega z_1 & z_1 & -z_1 & -\omega z_1  \\
	0 & 0 & 0 & 0  \\
	-z_3 & \omega z_3 & -\omega z_3 & z_3  \\
	\end{matrix} \right].
\end{equation}

Similarly, we can obtain the codebooks of the remaining users. In a downlink SCMA system, it is noted that the codewords simultaneously transmitted by the $J$ users are superimposed over the $K$ orthogonal resources. In total, there are $M^J$ superimposed codewords, which constitute a superimposed constellation $\vartheta$. The MED, a key indicator of the BER performance under AWGN channel, is obtained by calculating $M^J(M^J-1)/2$ mutual distances between the $M^J$ superimposed codewords, which can be expressed as
\begin{equation}
	\begin{aligned}
	\text{MED}=\min \left.\{ \left| \bm{x}\left( p \right)-\bm{x}\left( q \right) \right|,\forall \bm{x}\left( p \right),\bm{x}\left( q \right)\in\vartheta, \right. \\
	\left. \bm{x}\left( p \right)\ne \bm{x}\left( q \right) \right.\}, \\
	\end{aligned}
\end{equation}
where $\bm{x}(n)$ denotes the $n$-th superimposed codeword satisfying
\begin{equation}
	\bm{x}\left( n \right)=\sum\nolimits_{j=1}^{J}{{{\bm{x}}_{j}}\left( {{m}_{j}} \right)},
	m_j=1,2,\cdots,M.
\end{equation}

In a Rayleigh fading channel, on the other hand, the MPD of MC is an important performance indicator in high signal-noise ratio (SNR) ranges \cite{Boutros1996}. However, the MPD of each codebook may be different from that of MC due to the power imbalance among different users. Hence, it is of interest to calculate system MPD involving all the $J$ codebooks as follows:
\begin{footnotesize}
	\begin{equation}\label{MPD_equ}
	\text{MPD} = \min_{1\leq j \leq J}\left\{ \prod_{k=1,x_{j,p}^k\neq x_{j,q}^k}^K\left|x_{j,p}^k-x_{j,q}^k\right|,
	1\leq p<q\leq M \right\},
	\end{equation}
\end{footnotesize}
where $x_{j,m}^k$ denotes the $k$-th element of the $m$-th codeword associated to User $j$'s codebook.

Since MPD is only related to the non-zero dimensions of each codeword, without loss of generality, the non-zero dimensions (after removing all the zero rows) of a codebook can be expressed as
\begin{equation}
\bm{\Gamma}_{2\times M}=\left[\begin{matrix}
z_i & 0 \\
0 & z_j \\
\end{matrix}\right]\cdot \bm{C}_{2\times M}.
\end{equation}
By calculation, the MPD of constellation $\Gamma_{2\times M}$ can be expressed as
\begin{equation}
\text{MPD}(\bm{\Gamma}_{2\times M})=\frac{\sqrt{E_iE_j}}{E}\cdot \text{MPD}(\bm{C}_{2\times M}).
\end{equation}
where MPD consists of  both
$\frac{\sqrt{E_iE_j}}{E}$ and $\text{MPD}(\bm{C}_{2\times M})$. The first part is determined by the corresponding energy matrix, whereas the second part is determined by the MC. After careful analysis on the MC structure, it is found that the MPD of the MC $\bm{C}_{2\times N}$ can be obtained by calculating the MPD of the following three $2\times 2$ matrices $\bm{C}^{(1)}$, $\bm{C}^{(2)}$ and $\bm{C}^{(3)}$

\begin{equation}
\setlength{\arraycolsep}{1.2pt}
\bm{C}^{(1)}\!=\!\left[\begin{matrix}
+\omega_p & +\omega_q \\
+\omega_{\tilde{p}} & +\omega_{\tilde{q}} \\
\end{matrix}\right]\!,
\bm{C}^{(2)}\!=\!\left[\begin{matrix}
+\omega_p & +\omega_q \\
-\omega_{\tilde{p}} & +\omega_{\tilde{q}} \\
\end{matrix}\right]\!,
\bm{C}^{(3)}\!=\!\left[\begin{matrix}
-\omega_p & +\omega_q \\
-\omega_{\tilde{p}} & +\omega_{\tilde{q}} \\
\end{matrix}\right]\!,
\end{equation}
where $p+\tilde{p}=q+\tilde{q}=\frac{M}{2}+1$.

Let us assume $1\leq p\leq q\leq \frac{M}{2} (M>4)$ and $d=q-p$, then
\begin{equation}
\begin{split}
\text{MPD}(\bm{C}^{(1)})&=\vert\omega_q-\omega_p\vert\cdot\vert\omega_{\tilde{q}}-\omega_{\tilde{p}}\vert \\
&=d^2(\omega-1)^2 \\
&\geq 4(\omega-1)^2,
\end{split}
\end{equation}
where $d\geq 2$ is guaranteed to hold in $\bm{C}^{(1)}$ by the unique Star-QAM MC structure in (6).

We can further calculate the MPD of $\bm{C}^{(2)}$ as follows
\begin{equation}
\begin{split}
\text{MPD}(\bm{C}^{(2)})&=\vert\omega_q-\omega_p\vert\cdot\vert\omega_{\tilde{q}}+\omega_{\tilde{p}}\vert \\
&=d(\omega-1)\cdot \vert \omega_{\tilde{q}}+\omega_{\tilde{p}}\vert  \\
&\geq d(\omega-1)\cdot \vert \omega_{2}+\omega_{1}\vert \\
&=\omega^2-1,
\end{split}
\end{equation}
where $d\geq 1$ is also guaranteed to hold in $\bm{C}^{(2)}$ by the unique Star-QAM MC structure in (6).

Finally, we calculate the MPD of $\bm{C}^{(3)}$ as follows
\begin{equation}
\begin{split}
\text{MPD}(\bm{C}^{(3)})&=\vert\omega_q+\omega_p\vert\cdot\vert\omega_{\tilde{q}}+\omega_{\tilde{p}}\vert \\
&\geq \vert\omega_1+\omega_1\vert\cdot\vert\omega_{\frac{M}{2}}+\omega_{\frac{M}{2}}\vert   \\
&=(2M-4)(\omega-1)+4,
\end{split}
\end{equation}
where $d\geq 0$ is satisfied in  $\bm{C}^{(3)}$. Note that $(\omega_q+\omega_p)+(\omega_{\tilde{q}}+\omega_{\tilde{p}})=(\omega_q+\omega_{\tilde{q}})+(\omega_p+\omega_{\tilde{p}})=(\frac{M}{2}-1)(\omega-1)+1$ is a constant with fixed $M$ and $\omega$. When $p=q=1$ or $p=q=\frac{M}{2}$, the $\text{MPD}(\bm{C}^{(3)})$ is obtained.

Based on the above analysis, the MPD of MC $\bm{C}_{2\times N}$ can be written as \eqref{MPDMC}, which is shown in the top of next page. $I(M>4)$ is the indicator function which is equal to 1 if $M>4$, and 0 otherwise. The minimum value of $\frac{\sqrt{E_iE_j}}{E}$ can be obtained by energy matrix. Thus, the MPD of $\bm{\Gamma}_{2\times M}$ with energy matrices $\bm{E}_{4\times 6}$ and $\bm{E}_{5\times 10}$ can be expressed as follows

\newcounter{TempEqCnt}                         
\setcounter{TempEqCnt}{\value{equation}}       
\setcounter{equation}{19}
\newcounter{mytempeqncnt}
\begin{figure*}[!t]
	\normalsize
	\setcounter{mytempeqncnt}{\value{equation}}
	\begin{equation}
	\label{MPDMC}
	\setlength{\arraycolsep}{1.2pt}
	\text{MPD}(\bm{C}_{2\times M})=\begin{cases}
	4(\omega-1)^2,& 1<\omega\leq 1+\frac{5}{3}I(M>4); \\
	\omega^2-1,& 1+\frac{5}{3}I(M>4)< \omega\leq M-2+\sqrt{(M-3)^2+4}; \\
	(2M-4)(\omega-1)+4,& \omega > M-2+\sqrt{(M-3)^2+4}.
	\end{cases}
	\end{equation}
	\setcounter{equation}{\value{mytempeqncnt}}
	\hrulefill
	\vspace*{4pt}
\end{figure*}

\setcounter{TempEqCnt}{\value{equation}}       
\setcounter{equation}{20}
\begin{equation}
\begin{split}
\!\text{MPD}(\bm{\Gamma}_{2\times M})\!&\!=\!\min \left\{\frac{\sqrt{E_1E_3}}{E},\!\frac{E_2}{E}\right\}\!\cdot \! \text{MPD}(\bm{C}_{2\times M}), \\
\!\text{MPD}(\bm{\Gamma}_{2\times M})\!&\!=\!\min \left\{\frac{\sqrt{E_1E_4}}{E},\!\frac{\sqrt{E_2E_3}}{E}\right\}\!\cdot \!\text{MPD}(\bm{C}_{2\times M}). \\
\end{split}	
\end{equation}

%

Let $\bm{E}=[E_1,E_2,\cdots,E_{d_f}]^T$ and $\bm{\varphi}=[\varphi_1,\varphi_2,\cdots,\varphi_{d_f}]^T$. In order to design an excellent codebook suitable for both Gaussian and Rayleigh fading channels, we aim to maximize the MED while keeping the MPD greater or equal a certain threshold $\kappa$, i.e.,
\begin{subequations}
	\begin{alignat}{4}
	\label{opt}
	\max_{\bm{E},\bm{\varphi},\omega}\quad & \text{MED}\left(\bm{E},\bm{\varphi},\omega \right) \\
	\mbox{s.t.}\quad
	& \text{MPD}(\bm{\Gamma}_{2\times M})\geq\kappa,  \\
	& \sum_{i=1}^{d_f}E_i=\frac{MJ}{K}, \\
	& E_i>0,\quad \forall i=1,2,\cdots,d_f, \\
	& 0<\varphi_i<\pi, \quad \forall i=1,2,\cdots,d_f, \\
	& \omega > 1,
	\end{alignat}
\end{subequations}
where MED is a function of the parameters $\bm{E}$,$\bm{\varphi}$ and $\omega$ which can be expressed as $\text{MED}\left(\bm{E},\bm{\varphi},\omega\right)$. However, $\text{MPD}(\bm{\Gamma}_{2\times M})$ is only related to parameters $\bm{E}$ and $\omega$. The value of $\kappa$ should be judiciously selected in order to maintain a large MPD while maximizing the MED\footnote{In our optimization, $\kappa$ value corresponding to a known reference codebook may be set initially. However, it is observed that larger MED may not be obtained if $\kappa$ is too large.}. For fair comparison, the average energy of the codeword (a column of the codebook) is normalized to unity. Therefore, the total amount of energy of SCMA codebooks is $MJ$. There are a total of $K$ resource elements, thus yielding $\sum_{i=1}^{d_f}{E}_i={MJ}/{K}$.

MED is obtained by calculating the mutual distance among $M^J$ superimposed codewords, which is feasible for the codebook with $J=6$ and $K=4$. However, the calculation of MED may be very challenging when $M$ or $J$ are large\footnote{In comparison, as shown in (\ref{MPD_equ}), the system MPD is only dependent on the MPDs of $J$ codebooks which is much simpler to calculate.}. To solve this problem, a sub-optimal Monte Carlo search method is adopted in this paper. Firstly, the MED of $Q$ points which are randomly selected from superimposed constellation is calculated. Secondly, we repeat the above steps until the number of iteration reaches $t_{max}$. The final MED value after the $t_{max}$ iteration is specified as the MED of superimposed constellation. In this paper, we select $Q=5000$ and $t_{max}=20$.

Next, we employ GA\footnote{It is noted that our codebook optimization problem does not prevent the use of other heuristic algorithms. In particular, we have tested the performance of particle swam optimization (PSO), but found no advantage over GA solver. } from MATLAB Global Optimization Toolbox to solve \eqref{opt}. The obtained codebooks for $M=4,J=6,K=4$ and $M=16,J=10,K=5$ are shown in Appendix A. In principle, GA is a heuristic algorithm which carries out the search by simulating the natural evolution process. It starts from an initial solution called initial population. In each iteration, we select individuals according to their fitness in the optimization, and generate the next population through crossover and mutation operations. This process stops until it reaches the maximum number of iterations. The population size and the maximum number of iterations are set to 50 and 50, respectively. {In our optimization, we have randomly set the initial conditions and it can always converge to enhanced SCMA codebooks. Although different codebooks may be obtained after GA optimization, it is found that they exhibit more or less the same BER performances.}

\begin{figure}[htbp]
	\centering
	\includegraphics[width=3in]{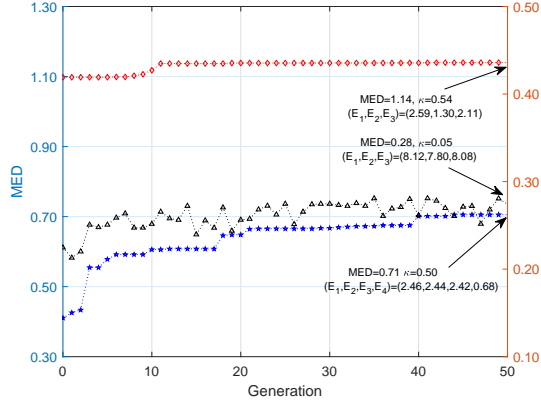} \\
	\caption{MED with increasing number of iterations in the proposed GA aided codebook optimization.\color{black}}
	\label{sim}
\end{figure}
\begin{figure}[htbp]
	\centering
	\includegraphics[width=3in]{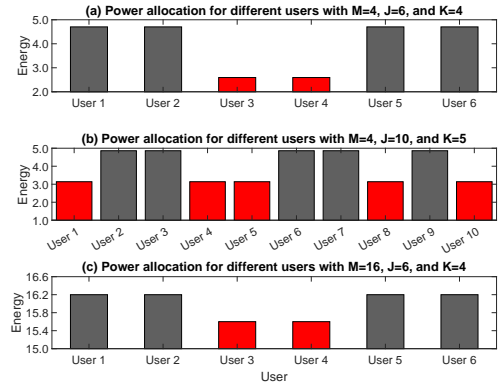} \\
	\caption{Energy distribution of different users in the proposed GA aided codebook optimization.\color{black}}
	\label{plotPower}
\end{figure}
The optimized MEDs with increasing number of iterations (up to 50) are shown in Fig. \ref{sim}, where the maximized MED is 1.14 for $K=4,J=6,M=4$. The optimum values of ${E}_1$, ${E}_2$ and ${E}_3$ are 2.59, 1.30 and 2.11, respectively, resulting in certain power imbalance among different users. Similar trends can be observed for $K=5, J=10, M=4$ and $K=4, J=6, M=16$, where the maximized MEDs are 0.71  and 0.28, respectively. It is noted that the values of $\kappa$ in the optimization for the three types of SCMA systems are set to be $0.54, 0.50,0.05$, respectively. The energy distributions of these optimized SCMA systems are shown in Fig. \ref{plotPower}.

\section{Numerical Comparison}
In this part, we compare our proposed codebooks with the GAM codebook \cite{Mheich2018}, Star-QAM codebook \cite{Yu2018}, Chen codebook \cite{Chen2020}, GA codebook\footnote{CS7 codebook is selected from \cite{Klimentyev2017} as it enjoys the largest MPD.} \cite{Klimentyev2017}, Huawei codebook \cite{Altera2013}, Xiao codebook \cite{Xiao2018} and Deka codebook \cite{Deka2020}.

Table \ref{Comparison of different codebooks} compares these codebooks with $M=4$,~$J=6$ and $K=4$. One can see that the power imbalance between different users is a distinctive feature of our proposed codebook. Comparing to other codebooks, our obtained codebook enjoys the largest MED by setting $\kappa=0.54$ which is equal to the MPD of the GA codebook. It is observed that the MPD of our proposed codebook is slightly higher than that of the GA codebook.

\begin{table}
	\centering
	\caption{Comparison of different codebooks with $M=4$,~$J=6$ and $K=4$}
	\begin{tabular}{|c | c| c | c |}
		\hline
			& \makecell[c]{power \\ imbalance} & MED & MPD \\
		\hline
		Star-QAM & No  & 0.90 & 0.72 \\
		\hline
		Xiao & No & 0.82 & 0.82 \\
		\hline
		GA & No & 1.01 & 0.54 \\
		\hline
		Chen & No & 1.07 & 0.78 \\
		\hline
		Deka & No & 0.94 & 0.48 \\
		\hline
		{Proposed} & {Yes} & {1.14} & {0.55} \\
		\hline
	\end{tabular}
	\label{Comparison of different codebooks}
\end{table}

\begin{figure}[htbp]
	\centering
	\includegraphics[width=3in]{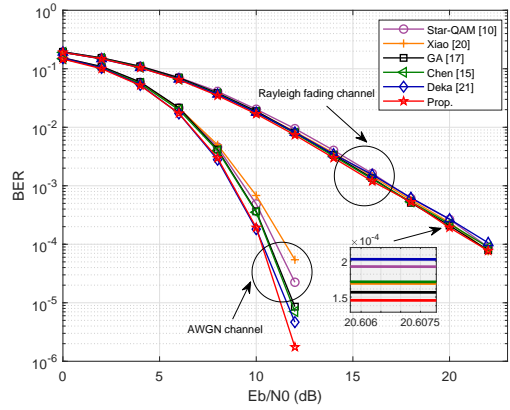} \\
	\caption{BER performance comparison of different codebooks for $M=4$,~$J=6$ and $K=4$ $(\lambda=150\%)$ under Gaussian channel and Rayleigh fading channel.}
	\label{BER_F4x6_M4}
\end{figure}

Fig. \ref{BER_F4x6_M4} shows the BER comparison of codebooks for $M=4$,~$J=6$ and $K=4$ $(\lambda=150\%)$. Unlike other codebooks, our obtained codebook enjoys higher MED. In Gaussian channel, our proposed codebook slightly outperforms the Deka codebook and achieves about 0.5 dB gain over the Chen codebook, 0.5 dB gain over the GA codebook, 0.8 dB gain over the Star-QAM codebook and 1.5 dB gain over the Xiao codebook at BER=$10^{-4}$, respectively. Owing to the inherent power-imbalance which reinforces the near-far effect among different interfering users, our proposed codebook achieves BERs which are slightly better than that of other codebooks in Rayleigh fading channel.

\begin{figure}[htbp]
	\centering
	\includegraphics[width=3in]{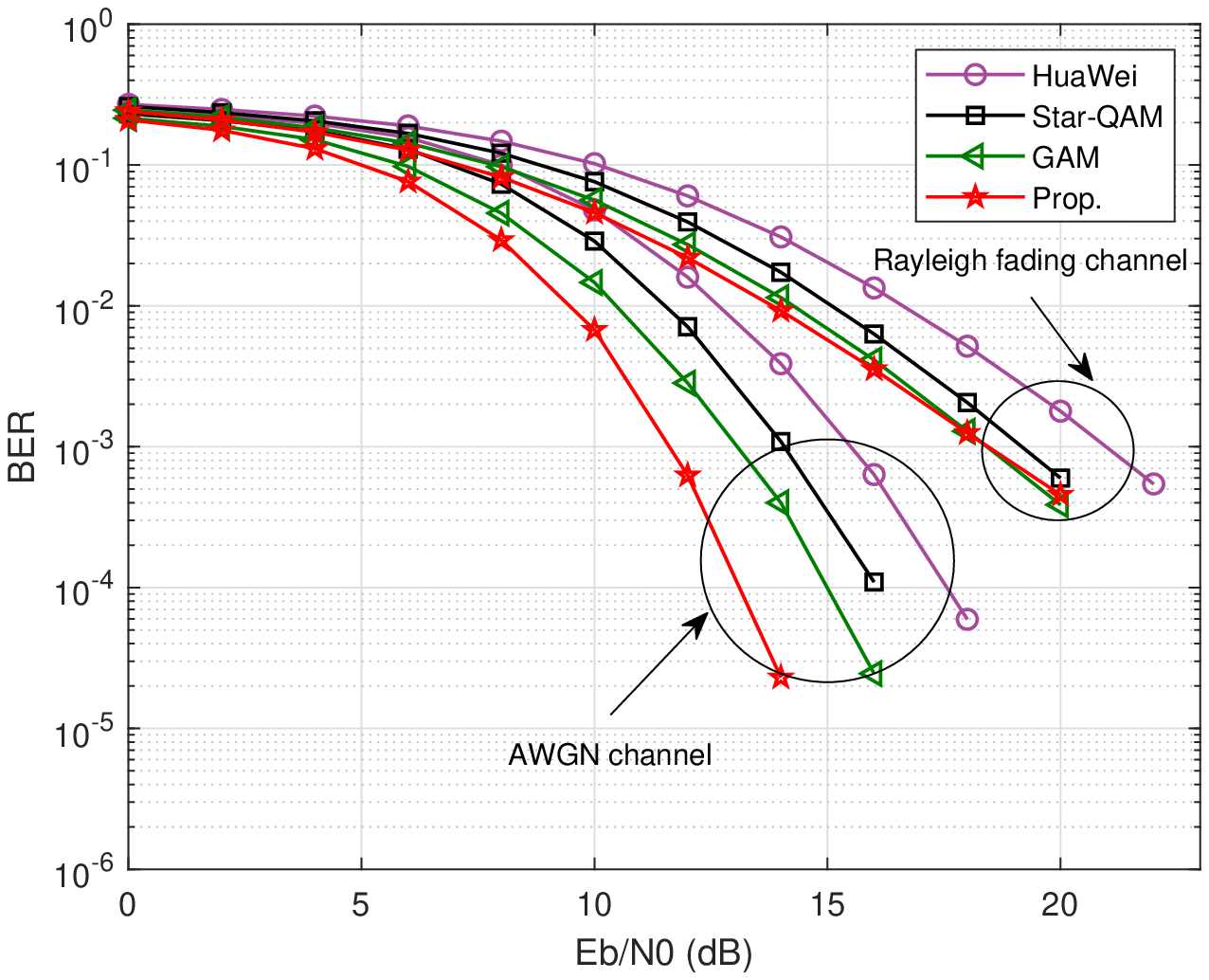} \\
	\caption{BER performance comparison of different codebooks for $M=4$,~$J=10$ and $K=5$ $(\lambda=200\%)$ under Gaussian channel and Rayleigh fading channel.\color{black}}
	\label{BER_F5x10_M4}
\end{figure}

Larger BER gains are observed for $M=4$,~$J=10$ and $K=5$ $(\lambda=200\%)$ in Fig. \ref{BER_F5x10_M4} due to the larger MED. In Gaussian channel, our codebook achieves about 1.9~dB gain over the GAM codebook, 3.0~dB gain over the Star-QAM codebook and 4.5~dB gain over the Huawei codebook at BER=$10^{-4}$, respectively. In Rayleigh fading channel, we obtain gains about 0.8 dB gain over the Star-QAM codebook and 2.6~dB gain over Huawei codebook at BER=$10^{-3}$.

The excellent error rate advantage of our proposed codebook can be seen more clearly for $M=16$,~$J=6$ and $K=4$ $(\lambda=150\%)$. In Gaussian channel, our proposed codebook achieves about 1.3 dB gain over the Huawei codebook, 1.5 dB gain over the GAM codebook and 1.6 dB gain over the Star-QAM codebook at BER=$10^{-3}$, respectively. In Rayleigh fading channel, our proposed codebook achieves about 1.5 dB gain over the Huawei codebook, 1.8 dB gain over the GAM codebook and 1.9 dB gain over the Star-QAM codebook at BER=$10^{-2}$, respectively.

\begin{figure}[htbp]
	\centering
	\includegraphics[width=3in]{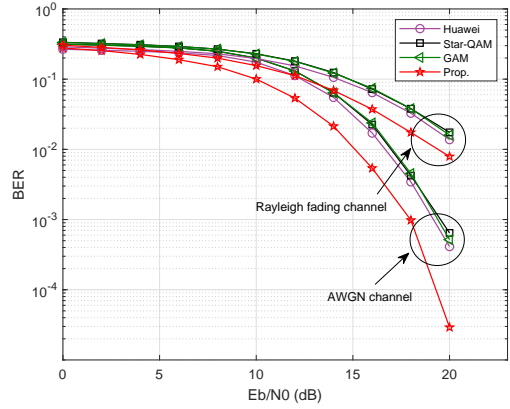} \\
	\caption{BER performance comparison of different codebooks for $M=16$,~$J=6$ and $K=4$ $(\lambda=150\%)$ under Gaussian channel and Rayleigh fading channel.}
	\label{BER_F4x6_M16}
\end{figure}

\begin{figure}[htbp]
	\centering
	\includegraphics[width=3in]{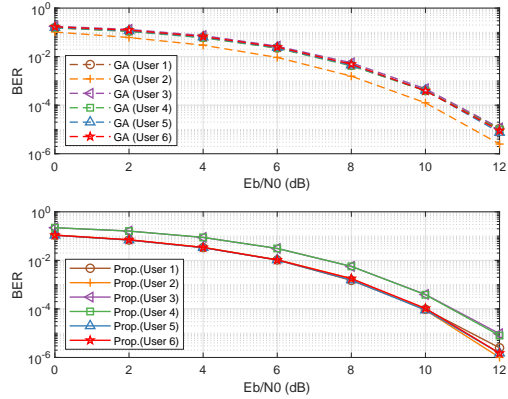} \\
	\caption{BER comparison for all users with $M=4$,~$J=6$ and $K=4$ $(\lambda=150\%)$ using the GA codebook and proposed codebook in Gaussian channel.}
	\label{UBER_F4x6_M4}
\end{figure}

Due to the power imbalance of the proposed codebook, the BER performances may vary among different  users. Fig. \ref{UBER_F4x6_M4} illustrates BER comparison for all users with $M=4$,~$J=6$ and $K=4$, including the GA codebook and our proposed codebook, in Gaussian channel. For the GA codebook, only User 2 performs better than the other five users with the performance gain of 0.8 dB at BER=$10^{-5}$. By contrast, for our proposed codebook, Users $1,2,5,6$ outperform the other two (which have the smaller power allocation) with performance gain of 0.9 dB at BER=$10^{-5}$. This explains the better average BER performance of our proposed codebook in Gaussian channel. The same trend can be observed for the Rayleigh fading channels, although the performance gains become smaller. 


%

\section{Conclusions}
In this paper, we have proposed a new class of SCMA codebook which displays power imbalance among different users. Simulation results have shown that our proposed power-imbalanced codebook offers substantial error rate gains over several known codebooks in Gaussian channel and Rayleigh fading channel. The noticeable gains stem from the fact that the power imbalance of the proposed codebooks helps to amplify the ``near-far effect'' which is useful for enhanced interference cancellation in MPA decoding.

{
\begin{appendices}	
	\section{}
	A. Power-imbalanced SCMA codebook for $M=4$,~$J=6$ and $K=4$ $(\lambda=150\%)$:
\begin{tiny}
	\begin{equation*}
	\mathbf{\bm{\chi}}_1=\left[\begin{matrix}
	\begin{smallmatrix}
	0 & 0 & 0 & 0 \\
	-0.2378 + 1.0684i & -0.0684 + 0.3074i & 0.0684 - 0.3074i & 0.2378 - 1.0684i \\
	0 & 0 & 0 & 0 \\
	-0.2840  & 0.9869 & -0.9869 & 0.2840 \\
	\end{smallmatrix}
	\end{matrix}\right],	
	\end{equation*}
\end{tiny}
\begin{tiny}
	\begin{equation*}
	\mathbf{\bm{\chi}}_2=\left[\begin{matrix}
	\begin{smallmatrix}
	-0.2378 + 1.0684i & -0.0684 + 0.3074i & 0.0684 - 0.3074i & 0.2378 - 1.0684i \\
	0 & 0 & 0 & 0 \\
	-0.2840  & 0.9869 & -0.9869 & 0.2840 \\
	0 & 0 & 0 & 0 \\
	\end{smallmatrix}
	\end{matrix}\right],
	\end{equation*}
\end{tiny}
\begin{tiny}
	\begin{equation*}
	\mathbf{\bm{\chi}}_3=\left[\begin{matrix}
	\begin{smallmatrix}
	0.6744 + 0.3794i & 0.1941 + 0.1092i & -0.1941 - 0.1092i & -0.6744 - 0.3794i \\
	-0.1941 - 0.1092i & 0.6744 + 0.3794i &  -0.6744 - 0.3794i & 0.1941 + 0.1092i \\
	0 & 0 & 0 & 0 \\
	0 & 0 & 0 & 0 \\
	\end{smallmatrix}
	\end{matrix}\right],
	\end{equation*}
\end{tiny}
\begin{tiny}
	\begin{equation*}
	\mathbf{\bm{\chi}}_4=\left[\begin{matrix}
	\begin{smallmatrix}
	0 & 0 & 0 & 0 \\
	0 & 0 & 0 & 0 \\
	0.6744 + 0.3794i & 0.1941 + 0.1092i & -0.1941 - 0.1092i & -0.6744 - 0.3794i \\
	-0.1941 - 0.1092i & 0.6744 + 0.3794i &  -0.6744 - 0.3794i & 0.1941 + 0.1092i \\
	\end{smallmatrix}
	\end{matrix}\right],
	\end{equation*}
\end{tiny}
\begin{tiny}
	\begin{equation*}
	\mathbf{\bm{\chi}}_5=\left[\begin{matrix}
	\begin{smallmatrix}
	0.9869 & 0.2840 & -0.2840 & -0.9869 \\
	0 & 0 & 0 & 0 \\
	0 & 0 & 0 & 0 \\
	0.0684 - 0.3074i & -0.2378 + 1.0684i & 0.2378 - 1.0684i & -0.0684 + 0.3074i \\
	\end{smallmatrix}
	\end{matrix}\right],
	\end{equation*}
\end{tiny}
\begin{tiny}
	\begin{equation*}
	\mathbf{\bm{\chi}}_6=\left[\begin{matrix}
	\begin{smallmatrix}
	0 & 0 & 0 & 0 \\
	0.9869 & 0.2840 & -0.2840 & -0.9869 \\
	0.0684 - 0.3074i & -0.2378 + 1.0684i & 0.2378 - 1.0684i & -0.0684 + 0.3074i \\
	0 & 0 & 0 & 0 \\
	\end{smallmatrix}
	\end{matrix}\right].
	\end{equation*}
\end{tiny}

B. Power-imbalanced SCMA codebook for $M=4$,~$J=10$ and $K=5$ $(\lambda=200\%)$:
\begin{tiny}
	\begin{equation*}
	\mathbf{\bm{\chi}}_1=\left[\begin{matrix}
	\begin{smallmatrix}
	-0.6927 + 0.7932i & -0.2312 + 0.2647i & 0.2312 - 0.2647i & 0.6927 - 0.7932i \\
	-0.1838 & 0.5509 & -0.5509 & 0.1838 \\
	0 & 0 & 0 & 0 \\
	0 & 0 & 0 & 0 \\
	0 & 0 & 0 & 0 \\
	\end{smallmatrix}
	\end{matrix}\right],	
	\end{equation*}
\end{tiny}
\begin{tiny}
	\begin{equation*}
	\mathbf{\bm{\chi}}_2=\left[\begin{matrix}
	\begin{smallmatrix}
	-0.5559 + 0.8876i & -0.1855 + 0.2962i & 0.1855 - 0.2962i & 0.5559 - 0.8876i \\
	0 & 0 & 0 & 0 \\
	-0.3038 - 0.1705i & 0.9104 + 0.5110i & -0.9104 - 0.5110i & 0.3038 + 0.1705i \\
	0 & 0 & 0 & 0 \\
	0 & 0 & 0 & 0 \\
	\end{smallmatrix}
	\end{matrix}\right],
	\end{equation*}
\end{tiny}
\begin{tiny}
	\begin{equation*}
	\mathbf{\bm{\chi}}_3=\left[\begin{matrix}
	\begin{smallmatrix}
	0.9104 + 0.5110i & 0.3038 + 0.1705i & -0.3038 - 0.1705i & -0.9104 - 0.5110i \\
	0 & 0 & 0 & 0 \\
	0 & 0 & 0 & 0 \\
	0.1855 - 0.2962i & -0.5559 + 0.8876i & 0.5559 - 0.8876i & -0.1855 + 0.2962i \\
	0 & 0 & 0 & 0 \\
	\end{smallmatrix}
	\end{matrix}\right],
	\end{equation*}
\end{tiny}
\begin{tiny}
	\begin{equation*}
	\mathbf{\bm{\chi}}_4=\left[\begin{matrix}
	\begin{smallmatrix}
	0.5509 & 0.1838 & -0.1838 & -0.5509 \\
	0 & 0 & 0 & 0 \\
	0 & 0 & 0 & 0 \\
	0 & 0 & 0 & 0 \\
	0.2312 - 0.2647i & -0.6927 + 0.7932i & 0.6927 - 0.7932i & -0.2312 + 0.2647i  \\
	\end{smallmatrix}
	\end{matrix}\right],
	\end{equation*}
\end{tiny}
\begin{tiny}
	\begin{equation*}
	\mathbf{\bm{\chi}}_5=\left[\begin{matrix}
	\begin{smallmatrix}
	0 & 0 & 0 & 0 \\
	-0.6927 + 0.7932i & -0.2312 + 0.2647i & 0.2312 - 0.2647i & 0.6927 - 0.7932i \\
	-0.1838 & 0.5509 & -0.5509 & 0.1838 \\
	0 & 0 & 0 & 0 \\
	0 & 0 & 0 & 0 \\
	\end{smallmatrix}
	\end{matrix}\right],
	\end{equation*}
\end{tiny}
\begin{tiny}
	\begin{equation*}
	\mathbf{\bm{\chi}}_6=\left[\begin{matrix}
	\begin{smallmatrix}
	0 & 0 & 0 & 0 \\
	-0.5559 + 0.8876i & -0.1855 + 0.2962i & 0.1855 - 0.2962i & 0.5559 - 0.8876i \\
	0 & 0 & 0 & 0 \\
	-0.3038 - 0.1705i & 0.9104 + 0.5110i & -0.9104 - 0.5110i & 0.3038 + 0.1705i \\
	0 & 0 & 0 & 0 \\
	\end{smallmatrix}
	\end{matrix}\right].
	\end{equation*}
\end{tiny}
\begin{tiny}
	\begin{equation*}
	\mathbf{\bm{\chi}}_7=\left[\begin{matrix}
	\begin{smallmatrix}
	0 & 0 & 0 & 0 \\
	0.9104 + 0.5110i & 0.3038 + 0.1705i & -0.3038 - 0.1705i & -0.9104 - 0.5110i \\
	0 & 0 & 0 & 0 \\
	0 & 0 & 0 & 0 \\
	0.1855 - 0.2962i & -0.5559 + 0.8876i & 0.5559 - 0.8876i & -0.1855 + 0.2962i \\
	\end{smallmatrix}
	\end{matrix}\right].
	\end{equation*}
\end{tiny}
\begin{tiny}
	\begin{equation*}
	\mathbf{\bm{\chi}}_8=\left[\begin{matrix}
	\begin{smallmatrix}
	0 & 0 & 0 & 0 \\
	0 & 0 & 0 & 0 \\
	-0.6927 + 0.7932i & -0.2312 + 0.2647i & 0.2312 - 0.2647i & 0.6927 - 0.7932i \\
	-0.1838 & 0.5509 & -0.5509 & 0.1838 \\
	0 & 0 & 0 & 0 \\
	\end{smallmatrix}
	\end{matrix}\right].
	\end{equation*}
\end{tiny}
\begin{tiny}
	\begin{equation*}
	\mathbf{\bm{\chi}}_9=\left[\begin{matrix}
	\begin{smallmatrix}
	0 & 0 & 0 & 0 \\
	0 & 0 & 0 & 0 \\
	-0.5559 + 0.8876i & -0.1855 + 0.2962i & 0.1855 - 0.2962i & 0.5559 - 0.8876i \\
	0 & 0 & 0 & 0 \\
	-0.3038 - 0.1705i & 0.9104 + 0.5110i & -0.9104 - 0.5110i & 0.3038 + 0.1705i \\
	\end{smallmatrix}
	\end{matrix}\right].
	\end{equation*}
\end{tiny}
\begin{tiny}
	\begin{equation*}
	\mathbf{\bm{\chi}}_{10}=\left[\begin{matrix}
	\begin{smallmatrix}
	0 & 0 & 0 & 0 \\
	0 & 0 & 0 & 0 \\
	0 & 0 & 0 & 0 \\
	-0.6927 + 0.7932i & -0.2312 + 0.2647i & 0.2312 - 0.2647i & 0.6927 - 0.7932i \\
	-0.1838 & 0.5509 & -0.5509 & 0.1838 \\
	\end{smallmatrix}
	\end{matrix}\right].
	\end{equation*}
\end{tiny}
\end{appendices}}

\end{document}